\documentclass[aps,letterpaper,twocolumn,showpacs]{revtex4}

\usepackage{graphicx} 
\usepackage{amsmath}

\begin{document}

\title{Fast Ground State Manipulation of Neutral Atoms in Microscopic Optical Traps}
\author{D.~D.~Yavuz, P.~B.~Kulatunga, E.~Urban, T.~A.~Johnson, N.~Proite, T.~Henage, T.~G.~Walker, and M.~Saffman}
\affiliation{Department of Physics, University of Wisconsin - Madison, 1150 University Avenue, Madison, WI, 53706}
\date{\today}

\begin{abstract}
We demonstrate Rabi flopping at MHz rates between ground hyperfine states of neutral $^{87}$Rb
atoms that are trapped in two micron sized optical traps. Using tightly focused laser beams we demonstrate high fidelity, site specific Rabi rotations with crosstalk on neighboring sites separated by  $8~\mu\rm m$  at the level of $10^{-3}$. Ramsey spectroscopy is used to  measure a dephasing time of $870~\mu\rm s$ which is $\approx$~5000 times longer than the time for a $\pi/2$ pulse. 
\end{abstract}
\maketitle

Over the last decade quantum computing has attracted much attention due to the possibility of solving certain problems much
faster than a classical computer \cite{chuang}. 
A number of different approaches are currently being pursued to build a
scalable quantum computer and significant progress has been made with trapped 
ions \cite{monroe}, nuclear magnetic
resonance \cite{chuang2}, single photons \cite{zeilinger}, and solid state josephson junctions \cite{yamamoto}.
Neutral atoms trapped by optical 
fields are also being studied intensively as a viable approach to demonstrating quantum logic.
Neutral atom approaches are attractive for a number of reasons starting with the availability of well developed techniques for laser cooling and trapping \cite{chu,heinzen} and the potential for scalability \cite{ertmer}. 
The  qubit basis states can be represented
by  ground state hyperfine levels which have long decoherence times and are therefore suitable for storing quantum information. 
The qubits can be rapidly initialized
and manipulated with near resonant optical fields
through optical pumping and stimulated Raman processes.
A number of protocols for two-qubit gates have
been proposed \cite{nagates} including ground state collisions, optically
induced short range dipole-dipole interactions,
and dipole-dipole interactions between highly excited Rydberg
levels \cite{cirac,protsenko,ryabtsev}. The Rydberg atom approach appears particularly attractive since it has the potential for 
achieving fast, MHz rate gates whose fidelity is only weakly dependent on the motional state of the atoms \cite{saffwalk2005}. 

We report here on progress towards demonstrating quantum logic operations using neutral atom qubits in optical traps. 
Recent achievements in neutral atom quantum computing include the implementation of a five qubit quantum register by Meschede and colleagues \cite{meschede1,meschede2} and subpoissonian loading of single atoms to nearby dipole traps by the Grangier group \cite{grangier1,grangier2}.
Advancing on these pioneering works, we demonstrate loading and ground state manipulation of neutral $^{87}$Rb atoms in two closely spaced microscopic optical
traps. By optically addressing each of these traps, we demonstrate two-photon Rabi flopping between ground hyperfine states
$|0\rangle \equiv |F=1,m_F=0 \rangle$ and $|1\rangle \equiv |F=2,m_F=0 \rangle$ at a rate of 1.36 MHz. This rate corresponds to a time period of 183~ns to
perform a $\pi/2$ Rabi rotation. The Rabi rotations are performed with negligible cross talk between the two traps: a $\pi$
rotation on one site causes less than $1.4 \times 10^{-3}  \pi$ rotation on the other site.  Using Ramsey spectroscopy, we
measure a dephasing time of 870
$\mu$s. To our knowledge, our results demonstrate the best figure of merit, (dephasing time)/(Rabi rotation time), achieved to date for
neutral atom quantum computing. Furthermore, our optical addressing scheme which uses acousto-optic modulators to spatially scan tightly focussed beams 
can be readily extended to address multiple qubit sites in a one- or two-dimensional array, which could form the basis for a scalable quantum logic device.  

We proceed with a detailed description of our experiment. As shown in Fig.~1, we start with a standard $\sigma^+$-
$\sigma^-$ 6-beam magneto optical trap (MOT) that is loaded from a background vapor in an ultrahigh vacuum, 16 cm diameter stainless steel chamber  \cite{wieman}. The MOT beams have a total intensity of 
$\approx$ 12 mW /cm$^2$ and are 12 MHz red detuned from the $F=2 \rightarrow F'=3$ cycling
transition. A repumping beam tuned to the $F=1 \rightarrow F'=2$ transition is superimposed with the MOT beams. The MOT produces an atom cloud with a density of $10^9 / \rm cm^3$.
\begin{figure}
\begin{center}
\includegraphics[width=9cm]{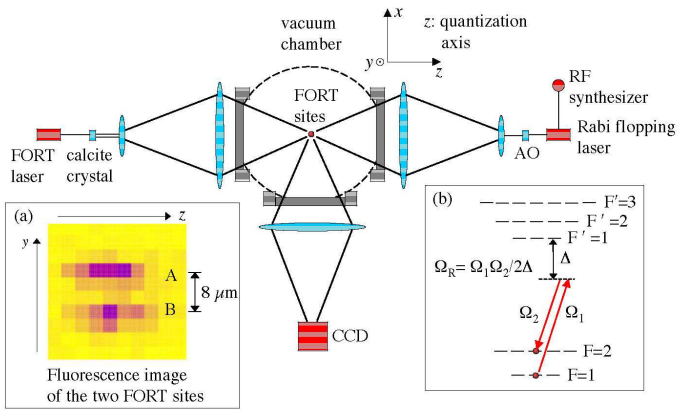}
\end{center}
\vskip 0in
\caption{(color online) The experimental setup. Two tightly focused beams that spatially overlap with a MOT 
form
closely spaced optical traps. The MOT beams are not shown for simplicity. Inset~(a) shows a fluorescence image of the
atoms that are trapped in the two micron sized FORTs (termed sites A and B  respectively).  Inset~(b) shows the relevant energy level structure of $^{87}$Rb. Two laser beams whose
frequency difference equals the hyperfine transition frequency implement Rabi rotations on the qubit states. The two laser beams are obtained by microwave modulating a diode laser. The Rabi
rotation beams are steered to either site~A or site~B with the use of an acousto optic (AO) modulator. 
}
\label{fig1}
\end{figure}

Two tightly focused beams that spatially overlap with the MOT cloud form two far-off resonant traps (FORT). 
The FORT
beams are obtained from a  diode laser at a wavelength of 1010~nm. The
output of this laser is split into two beams with the use of a birefringent calcite crystal. The
two spatially separated beams are then imaged into the center of the chamber with a custom designed lens system (NA 0.38). The resulting 80~mW FORT beams are focused to a near diffraction limited waist of $w_f=2.7~\mu\rm m$. The separation between the two FORT sites at the focus is $d=8~\mu\rm m$. The linearly polarized FORT beams form
$\approx$~1~mK deep potential wells at the focus. We typically
load about 10 atoms from the MOT into each of the FORT sites. The temperature of the atoms that are trapped in the FORTs is
measured to be $70~\mu\rm K$. The $1/e$ lifetime of the atoms in the FORT sites is 780~ms, limited by collisions with the
background vapor. 

In Fig.~1, inset~(a), we show a false-color fluorescence image of the atoms that are trapped in the two FORT sites 
A and B. The image is taken with an electron-multiplying CCD camera and the fluorescence from the atoms is collected with
a  custom lens system (NA 0.57) that is verified to have a resolution of 3~$\mu$m at the position
of the FORT sites. The image of Fig.~1 and the rest of the data that is
presented in this letter is taken using the following loading and measurement cycle. After loading the MOT for several
seconds, we reduce the intensity of the MOT beams by a factor of two and increase the detuning 
to 18 MHz for a period of 100~ms thereby cooling the atoms to a temperature of
30~$\mu K$ and loading the two FORT sites.  We turn off the MOT and repumping beams for 100 ms and let the atoms that are trapped in the MOT diffuse out of the viewing region. We then apply Rabi/Ramsey pulses as desired and probe the $F=2$ atoms that are trapped in the FORTs for 10~ms with the MOT beams
attenuated to an intensity of $100~\mu\rm W/cm^2$. The $F=1$ atoms are completely dark to the probing light. The FORT laser AC stark shifts the cycling transition. The mean shift for different Zeeman $m$-level transitions is measured to be 40~MHz to the blue. The frequency of the  probing beam is tuned to compensate for this mean shift. After 10~ms of probing, the atoms boil out of the trap
and are lost \cite{explanation1}. Inset (a) in Fig.~1 is an average of nine images (each with a 10~ms exposure time). Using
known parameters of the probing beam, the collection optics and the CCD camera sensitivity, we estimate 
single atom photoelectron rates of $2100$~s$^{-1}$ from atoms trapped in the FORT sites. Given this estimate, the image in Fig. 1  corresponds to $\approx$~10 atoms per site. 

As shown in  inset~(b) of Fig.~1, two-photon Rabi rotations between states $|0\rangle$ and $|1\rangle$ are
performed with two laser beams whose frequency difference equals the hyperfine transition frequency of 6,834,683~kHz. The two
beams are obtained by modulating the current of a single diode laser at half the transition frequency.
This modulation produces two sidebands with the desired frequency separation. The carrier is then removed with the use of a
filtering cavity with a finesse of 50. The two sidebands pass through an acousto-optic (AO) modulator and are focused to a near-diffraction
limited waist of $w=4.1~\mu\rm m$ in the chamber where they overlap with  FORT sites A or B. We individually address the
two FORT sites by changing the acoustic wave frequency (and thereby the diffraction angle) of the AO modulator. The frequency shift of the individual beams caused by the AO modulator is not of importance due to the two-photon nature of the stimulated Raman process. The total power 
in the two sidebands is $45~\mu\rm W$ and the detuning from the excited state is $\Delta =-2 \pi \times 41$~GHz. The polarization of the sidebands  is identical and is circular with respect to the quantization axis $z$.

We proceed with a detailed discussion of Rabi rotations on ground hyperfine states. With the Rabi
frequencies of the individual beams denoted as
$\Omega_1$ and $\Omega_2$, the two-photon driving Rabi frequency between the logical qubit states $|0\rangle$ and $|1\rangle$ is $\Omega_R =
\Omega_1
\Omega_2/2 \Delta
$. Here, $\Delta $ is much larger than the decay rate of
the excited state. In Fig.~2, we demonstrate fast Rabi rotations on one of the FORT sites with negligible cross-talk to the other site. The initial state is selected by turning off 
 the hyperfine repumping beam for a duration of 8~ms at the end of the FORT loading cycle, and thereby
optically pumping all the atoms into the $F=1$ Zeeman states. After this optical pumping we apply a bias magnetic field of 10.7 G along
the quantization axis to separate out different Zeeman $m$-level transition frequencies thereby isolating $m=0$ atoms. We then apply a two-photon Rabi
pulse of variable duration and probe the percentage of atoms that make a transition from $|0 \rangle$ to $|1 \rangle$. Plot~(a) shows the Rabi flopping of atoms in FORT
site~A. Each data point is an average of 12 experimental runs. The normalization of the vertical axis is obtained by optically pumping all the atoms into $F=2$ and observing the total fluorescence. The uncertainty of this normalization is $\pm 10 \%$.  We
observe a sinusoidal variation with high contrast \cite{explanation2}. The solid line is a sinusoidal fit and yields a two-photon Rabi frequency of
$\Omega_R = 2 \pi \times1.36~\rm MHz$ \cite{explanation3}. Considering that a single qubit Hadamard gate requires a $\pi/2$ rotation, this rate
corresponds to a single qubit manipulation time of $\pi/2\Omega_R$=183~ns. This is a factor of 44 faster than what has
previously been achieved in neutral atom quantum computing \cite{meschede1}. 

\begin{figure}
\begin{center}
\includegraphics[width=9.5cm]{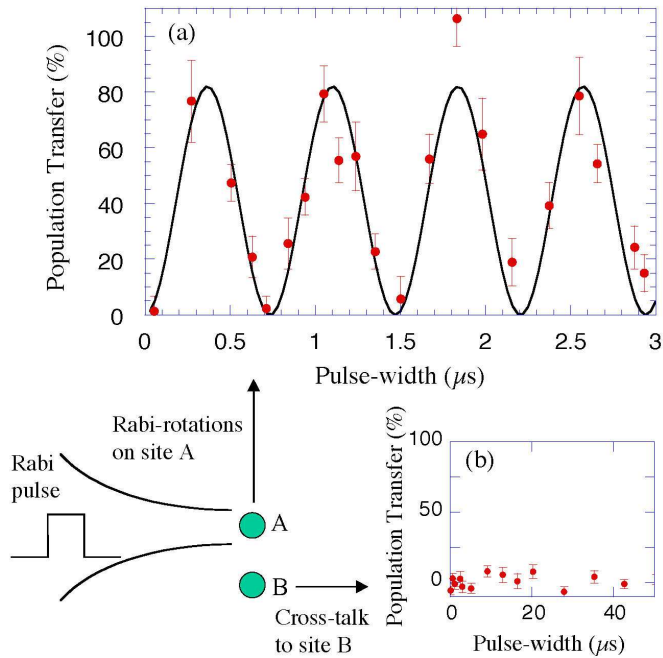}
\end{center}
\vskip 0in
\caption{(color online) Fast rabi rotations on one site with negligible cross-talk to the other site. In plot~(a), we start with all the atoms
in state $|0\rangle$, and measure the fraction of atoms in state $|1\rangle$ as a function of the duration of the two-photon
pulse. We observe Rabi flopping between two qubit states at a rate of $\Omega_R=2 \pi \times 1.36$~MHz. This rate corresponds
to a $\pi/2$ rotation time of 183~ns (see text for details). Plot~(b) shows that the cross-talk to the other site
is negligible and demonstrates our ability to individually address the two sites. 
}
\label{fig1}
\end{figure}

The performance of single qubit addressing is a key benchmark for quantum computing. In ion trap quantum computing, addressing errors of $\approx 10^{-2}$ have been demonstrated using focused beams \cite{blatt} and precise control of the micromotion of the ions \cite{turchette}. Plot~(b) in Fig.~2 shows the population transfer at FORT site~B, while Rabi flopping beam is aligned  to site~A. We do not see appreciable
excitation for pulse-widths as large as 43~$\mu$s. With our detection sensitivity of $\pi/6$ rotation, this implies that the  crosstalk (ratio of Rabi frequencies: $\Omega_{R}({\rm site~ B})/\Omega_{R}({\rm site~ A})$) is  less than $1.4\times10^{-3}$.  This upper bound on the crosstalk is only a few times higher than the theoretical value of  $e^{-2d^2/w^2}$ which evaluates to $ 4.9 \times 10^{-4}$ for our experimental parameters. With the help of the AO modulator, we can switch the Rabi flopping beam to address FORT~site~B instead of
FORT~site~A. For this case, we have verified that we repeat the results of Fig~2, with site~A and site~B interchanged. To eliminate detrimental effects of the resolution of imaging optics, the data of Fig.~2 is taken one site at a time. An
important advantage of the optical addressing scheme is that the amount of cross talk is independent of the speed of the
single-qubit operation. This is in contrast to the magnetic addressing scheme where higher speeds would require larger field gradients\cite{meschede1}. For example, a 1 MHz Rabi flopping rate with pulse area cross talk for 8~$\mu$m separated sites at the 10$^{-3}$ level would require a B-field gradient of greater than $10~\rm T/cm$ using $m=\pm1$ Zeeman states.

\begin{figure}
\begin{center}
\includegraphics[width=8.cm]{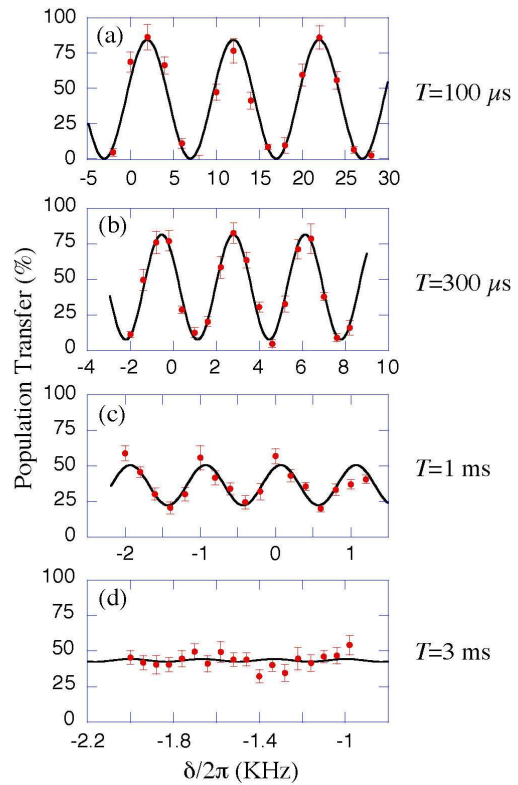}
\end{center}
\vskip 0in
\caption{(color online) Ramsey spectroscopy on the $|0\rangle \rightarrow |1\rangle$ hyperfine transition. We apply two $\pi /2$ pulses with a
delay $T$ and measure the fraction of atoms in state $|1\rangle$ as a function of the two-photon detuning $\delta$. }
\label{fig1}
\end{figure}

We next proceed with our measurements of the decoherence time of the qubit states. For this purpose, we use
Ramsey's method of separated oscillatory fields \cite{ramsey,kasevich}. With all atoms starting in state $|0 \rangle$, we apply
two $\pi /2$ pulses that are separated by a time $T$. We then measure the fraction of atoms that make a transition to
state $|1\rangle$ as a function of the two photon detuning $\delta$. The contrast of the fringe patterns are expected to decay exponentially with a time constant $T_2$ which is the dephasing time of the $|0\rangle$ to $|1\rangle$
hyperfine transition. In Fig.~3, plots (a) to (d) show the result of this measurement for $T$=100~$\mu$s, 300~$\mu$s, 1~ms, and
3~ms respectively. Each data point is again an average of 12 experimental runs. The solid line in each plot is a sinusoidal
fit with an offset. As expected we observe a fringe pattern with a reduced contrast as $T$ becomes larger. In Fig.~4, we plot
the contrast of the fringes in Fig.~3, as a function of $T$. The best exponential fit to the data points yields a dephasing time
of $T_2=870$~$\mu$s. This gives a figure of merit of (dephasing~time)/($\pi/2$ Rabi~rotation~time)~=~($870$~ $\mu$s)/($183$~ns)=4750.

\begin{figure}
\begin{center}

\includegraphics[width=8.cm]{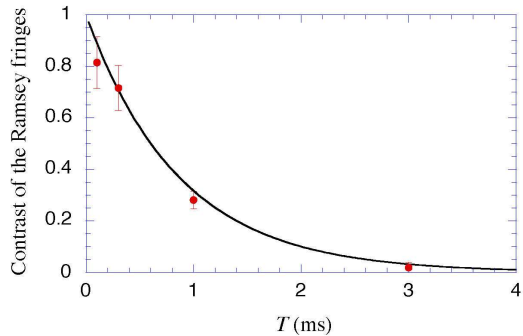}
\end{center}
\vskip 0in
\caption{(color online) The contrast of the Ramsey fringes of Fig.~3 as a function of the time delay between two pulses, $T$. The solid line is an exponential fit to the data. 
}
\label{fig1}
\end{figure}

In conclusion, we have demonstrated site specific ground state manipulation at MHz rates and long decoherence times of hyperfine transitions of neutral atoms in two nearby dipole traps. In the future, it will be relatively straightforward to increase the power of the Rabi
flopping beams by several orders of magnitude and thereby obtain manipulation rates in the GHz range. Combining such fast
one-qubit gates with  two-qubit Rydberg gates \cite{cirac} may provide a powerful building block for a scalable
quantum computer. 

We would like to thank Antoine Browaeys and Philippe Grangier for helpful discussions. This work was
supported by the U. S. Army Research Office under contract number DAAD19-02-1-0083 and NSF grant PHY-0205236.


\begin{references}
\bibitem{explanation1} Our results differ from the results of other groups where probing of atoms for several seconds without significant heating has been reported (see for example Ref. \cite{grangier1}). One possible explanation is the large AC stark shift of the cycling transition due to the FORT laser at 1010~nm. We plan to investigate this further in the future.
\bibitem{explanation2} The stark shift of the hyperfine transition due to the Rabi rotation beams is calculated to be -$0.33 \times \Omega_R/2\pi$=-449 kHz. The quadratic Zeeman shift of the transition from the 10.7 G bias field is 66 kHz. Because of these two effects, the two-photon excitation is slightly off-resonant which may partially explain the 82\% contrast of Fig.~2(a).
\bibitem{explanation3} The expected Rabi frequency for our experimental parameters is $\Omega_R = 2 \pi \times3.75~\rm MHz$. The discrepancy may be due to misalignment of the Rabi rotation beams from the addressed FORT site.
\end{references}
\end {document}